\documentclass[notitlepage,showkeys,onecolumn,11pt,times,superscriptaddress, a4paper, pre, nofootinbib]{revtex4-2}
\usepackage[english]{babel}
\usepackage{stix}
\usepackage[T1]{fontenc}
\usepackage{graphicx}
\usepackage{bm} 
\usepackage[colorlinks=true,
            linkcolor=blue,
            urlcolor=blue,
            citecolor=blue]{hyperref}
\usepackage[mathlines]{lineno} 
\usepackage{amsmath}
\usepackage{physics}
\usepackage{booktabs}
\usepackage{geometry}
\newcommand*\dif{\mathop{}\!\mathrm{d}}

\begin{document}

    \title{On the dynamical and statistical properties of a quartic mean--field Hamiltonian model}

    \author{Matheus Rolim Sales}
    \email{rolim.sales.m@gmail.com}
    \affiliation{São Paulo State University (UNESP), Institute of Geosciences and Exact Sciences, 13506-900, Rio Claro, SP, Brazil}
    \affiliation{University of Essex, School of Mathematics, Statistics and Actuarial Science, CO4 3SQ, Wivenhoe Park, Colchester, United Kingdom}
    \author{Edson Denis Leonel}
    \affiliation{São Paulo State University (UNESP), Institute of Geosciences and Exact Sciences, 13506-900, Rio Claro, SP, Brazil}
    \author{Chris G. Antonopoulos}
    \affiliation{University of Essex, School of Mathematics, Statistics and Actuarial Science, CO4 3SQ, Wivenhoe Park, Colchester, United Kingdom}

    \begin{abstract}
        Mean--field systems provide a natural framework in which collective effects persist as the number of degrees of freedom $N$ increases, raising fundamental questions about the emergence of integrability and the nature of chaos in large but finite systems. We investigate the dynamical and statistical properties of a quartic mean-field Hamiltonian model, with particular emphasis on the relation between the thermodynamic limit and finite--size chaotic dynamics. We first analyze the thermodynamic limit of the model within the Vlasov collisionless framework and derive the corresponding self--consistent single--particle description. We identify the conditions under which the mean-field dynamics becomes effectively autonomous and show numerically that fluctuations of the relevant intensive quantities vanish algebraically with $N$, supporting the emergence of integrability as $N \to \infty$. We then study the finite--$N$ dynamics by computing the largest Lyapunov exponent over an exceptionally wide range of $N$, spanning several orders of magnitude. We find that the largest Lyapunov exponent decays algebraically with $N$, consistently with the suppression of chaos in the thermodynamic limit for mean--field Hamiltonian models. Using tools from non--extensive statistical mechanics, we further analyze the time evolution of the entropic index $q$ and demonstrate that, although transient values $q > 1$ may appear at intermediate times, $q$ systematically converges to unity as the observation time increases. This behavior indicates that the finite--$N$ dynamics is strongly chaotic in the asymptotic regime and that previously reported $q > 1$ values for the present models originate from finite-time effects rather than from a persistent weakly chaotic phase. 
    \end{abstract}
        
    \date{\today}
    \maketitle

    \section{Introduction}
    \label{sec:introduction}
    
    Mean--field and long--range interacting Hamiltonian systems provide a paradigmatic setting in which collective effects persist in the thermodynamic limit and strongly influence both microscopic dynamics and macroscopic statistical behavior~\cite{Dauxois2002,Campa2009,Bouchet2010,Gupta2017}. Because each degree of freedom interacts with all others, typically under a Kac scaling, these models exhibit features that are absent in short-range systems, including nontrivial finite-size effects, slow relaxation, and an interplay between dynamics and statistical mechanics that remains an active topic of research~\cite{Anteneodo1998,Latora1998,Latora1999,Cirto2014,Dauxois2002,Campa2009,Bouchet2010,Manos2011, Christodoulidi2014,Christodoulidi2016,Christodoulidi2018,Bagchi2016,Bagchi2017,Gupta2017,Bagchi2017b,Iubini2018,Miloshevich2017}. A central question in this context concerns the nature of chaos as the number of degrees of freedom, $N$, increases: while finite systems can display strong sensitivity to initial conditions~\cite{Anteneodo1998,Bagchi2016}, the thermodynamic-limit description may become effectively regular, and understanding how these two pictures connect requires a careful combined dynamical and statistical analysis~\cite{Latora1998,Latora1999,Firpo1998,Casetti2000,Anteneodo2001,Firpo2001,Manos2011,Ginelli2011, Bagchi2018}.
    
    For mean--field Hamiltonians, the limit $N\to\infty$ admits a continuum description in terms of a one--particle distribution evolving under a self--consistent field. This thermodynamic--limit framework is rigorously justified for broad classes of mean--field interactions and initial conditions and is naturally expressed through the Vlasov collisionless equation~\cite{Braun1977,Dobrushin1979,Latora1999,Antoniazzi2007}. In this limit, the microscopic dynamics of each particle is effectively governed by the collective mean field generated by the full system, leading to an effective single--particle description. In contrast, for any finite $N$, discreteness effects generate fluctuations around the limiting mean field. These finite--size fluctuations are expected to vanish as $N$ increases, yet they still produce measurable chaoticity and control the scaling of the largest Lyapunov exponent with $N$~\cite{Casetti1995,Casetti1996, Latora1998,Firpo1998,Anteneodo2001,Firpo2001}.
    
    A recent work on a quartic mean--field model has reported an algebraic decay of the largest Lyapunov exponent with $N$ and interpreted this behavior as indicative of weak chaos, supported by a transient $q > 1$ entropic index obtained from $q$--Gaussian fits of dynamical probability density functions~\cite{Christodoulidi2025}. Such interpretations naturally connect to non--extensive statistical mechanics approaches, where deviations from Boltzmann--Gibbs statistics and $q$--Gaussian distributions are often discussed in relation to long--lived correlations and quasi-stationary dynamical regimes~\cite{Tsallis2009,Moyano2006,Antonopoulos2011b,Bountis2012}. At the same time, the estimation of $q$ from finite-time data is known to be sensitive to statistical sampling and to the slow convergence of dynamical observables, making it essential to distinguish genuine long--lived non-Boltzmann--Gibbs behavior from finite-time and finite-size effects~\cite{Gerakopoulos2008,Ruiz2017,Ruiz2017b,Tirnakli2016,Zulkarnain2026}.
    
    In this work, we revisit the quartic mean--field Hamiltonian model~\cite{Christodoulidi2025} and provide a dynamical and statistical characterization across both the thermodynamic--limit description and the finite--$N$ dynamics. We first derive the thermodynamic--limit mean--field framework and identify the conditions under which the corresponding single--particle dynamics becomes effectively autonomous. We then show that, under these conditions, the thermodynamic--limit dynamics reduces to an integrable one--degree--of--freedom Hamiltonian system. We present numerical evidence that the fluctuations of the relevant intensive quantities vanish algebraically with $N$, supporting the emergence of integrability as $N\to\infty$. Next, we compute the largest Lyapunov exponent over a very broad range of numbers of degrees of freedom and confirm its systematic decay with $N$, consistent with the suppression of chaos in the large-$N$ limit~\cite{Anteneodo1998, Firpo1998, Latora1998, Latora1999, Firpo2001, Anteneodo2001, Dauxois2002, Ginelli2011, Manos2011, Christodoulidi2014, Christodoulidi2016, Christodoulidi2018, Christodoulidi2025}. Finally, guided by non--extensive statistical mechanics concepts~\cite{Tsallis2009}, we study the time evolution of the entropic index $q(t)$ together with fit-quality diagnostics and show that $q(t)\to 1$ as the observation time increases, indicating that the finite--$N$ dynamics is strongly chaotic asymptotically and that transient $q>1$ values originate from finite-time effects rather than from a persistent weakly chaotic regime~\cite{Moyano2006,Antonopoulos2011b,Bountis2012, Christodoulidi2025}.

    This paper is organized as follows: In Sec.~\ref{sec:model}, we introduce the quartic mean--field Hamiltonian model, which is the main object of study of this work, and derive its thermodynamic-limit description in terms of the Vlasov collisionless equation. We also discuss the conditions for integrability in this limit and provide numerical evidence showing that the model becomes integrable as $N\to\infty$. In Sec.~\ref{sec:finiteNdynamics}, we investigate the dynamical and statistical properties of the system at finite $N$. We compute the largest Lyapunov exponent (LLE) and show that it decreases with increasing number of degrees of freedom, vanishing in the thermodynamic limit, consistently with the analysis presented in Sec.~\ref{sec:model}. Guided by concepts from non--extensive statistical mechanics~\cite{Tsallis2009}, we further demonstrate that the finite--$N$ dynamics is strongly chaotic, rather than weakly chaotic as suggested in a recent study~\cite{Christodoulidi2025}. Finally, Sec.~\ref{sec:concl} contains our conclusions and final remarks.

    \section{The quartic mean--field model and its thermodynamic limit}
    \label{sec:model}

    \subsection{The model}
    \label{subsec:model}

    We study an autonomous mean--field Hamiltonian system of $N$ globally coupled sites, described by the Hamiltonian
    \begin{equation}
        \label{eq:hamiltonian}
        H(\mathbf{p}, \mathbf{x}) = \sum_{n = 1}^N\left[\frac{p_n^2}{2} + V(x_n) + \frac{1}{2\tilde N}\sum_{m = 1}^N W(x_m - x_n)\right] \equiv E,
    \end{equation}
    where $x_n$ and $p_n$ are the generalized coordinates and momenta, respectively, $E$ is the total mechanical energy, $\mathbf{p} = (p_1,\ldots,p_N)$, $\mathbf{x} = (x_1,\ldots,x_N)$, and $\tilde N = N-1$.
    The factor $1/\tilde N$ is the Kac normalization and guarantees that the total energy $H$ is extensive, i.e., $H \propto N$, in the thermodynamic limit. We consider quartic on-site and interaction potentials~\cite{Christodoulidi2025},
    \begin{equation}
        \begin{aligned}
        V(x_n) &= \frac{x_n^2}{2} + \frac{x_n^4}{4},\\
        W(x_m - x_n) &= \frac{1}{4}(x_m - x_n)^4 .
        \end{aligned}
    \end{equation}
    The equations of motion are
    \begin{equation}
        \label{eq:eom}
        \dot{x}_n = p_n,\quad
        \dot{p}_n = -\left[x_n + x_n^3 + \frac{1}{\tilde N}\sum_{m = 1}^N (x_n - x_m)^3\right] .
    \end{equation}
    To integrate the equations of motion we use the fourth-order Yoshida symplectic integrator~\cite{Yoshida1990} with time step $\Delta t = 0.01$.
    For all values of the specific energy $\varepsilon = E/N$ considered in this paper, the relative energy error satisfies
    \begin{equation*}
        E_r = \left|\frac{E(t)-E(0)}{E(0)}\right| \le 10^{-5}.
    \end{equation*}

    \subsection{The Vlasov limit}
    \label{subsec:vlasovlimit}
    
    Expanding the cubic term in the second equation in Eq.~\eqref{eq:eom} and summing over $m$, we get:
    \begin{equation*}
        \sum_{m = 1}^N(x_n - x_m)^3 = N x_n^3 - 3 x_n^2\sum_{m=1}^Nx_m + 3 x_n \sum_{m=1}^Nx_m^2 - \sum_{m=1}^Nx_m^3.
    \end{equation*}
    Let us then define the intensive (microscopic) moments of $x_m$ as
    \begin{equation}
        \label{eq:moments}
        M_k = \frac{1}{N}\sum_{m=1}^Nx_m^k,
    \end{equation}
    so that
    \begin{equation*}
        \sum_{m = 1}^N(x_n - x_m)^3 = N x_n^3 - 3 Nx_n^2M_1 + 3 Nx_n M_2 - NM_3.
    \end{equation*}
    Substituting back into the equations of motion given by Eq.~\eqref{eq:eom}, we obtain
    \begin{equation*}
        \dot{x}_n = p_n,\quad\dot{p}_n = - \left[x_n + x_n^3 + \frac{N}{\tilde N}\left(x_n^3 - 3 x_n^2M_1 + 3 x_n M_2 - M_3\right)\right].
    \end{equation*}
    Since $N/\tilde N = 1 + 1 / \tilde N$, i.e., a constant plus a term that goes to zero as $N \rightarrow \infty$, we can write
    \begin{equation}
        \label{eq:eom2}
        \dot{x}_n = p_n,\quad\dot{p}_n = -\left(x_n + 2x_n^3 - 3 x_n^2M_1 + 3 x_n M_2 - M_3\right) + \mathcal{O}(N^{-1}).
    \end{equation}
    Thus, the dependence of each particle on the rest of the system is through the intensive moments $M_k$. Let us define a probability distribution in phase space that represents how the particles are arranged at each instant of time:
    \begin{equation}
        \label{eq:particlesdistr}
        \rho_N(x, p, t) = \frac{1}{N}\sum_{m = 1}^N\delta(x - x_m(t))\delta(p - p_m(t)).
    \end{equation}
    In this way, any average over particles becomes an integral with respect to $\rho_N$
    \begin{equation}
        \label{eq:moments2}
        M_k(t) = \frac{1}{N}\sum_{m = 1}^Nx_m^k(t) = \int x^k(t)\rho_N(x, p, t)\dif{x}\dif{p}.
    \end{equation}
    By substituting Eq.~\eqref{eq:particlesdistr} into the integral in Eq.~\eqref{eq:moments2}, and using the filter property of the Dirac delta function, we recover the intensive moments defined in Eq.~\eqref{eq:moments}.

    When $N$ is small, the distribution $\rho_N$ is spiky and irregular. As $N$ increases toward $N\to\infty$, $\rho_N$ becomes increasingly dense in phase space, and particle averages converge to well-defined macroscopic limits. This convergence is ensured by the smoothness of the interactions and by the absence of strong correlations in the initial conditions~\cite{Braun1977,Dobrushin1979}. Moreover, because each particle is connected to all others through many links whose individual strength is weak due to the Kac scaling $1/N$, and because the Hamiltonian is invariant under permutations of particle indices, correlations between any finite number of particles become negligible in the thermodynamic limit, a property known as propagation of chaos~\cite{Sznitman1991,DelMoral2004}. In this regime, particle averages become self-averaging quantities and converge to ensemble averages defined with respect to a smooth microscopic one--particle distribution $f(x,p,t)$~\cite{Braun1977, Antoniazzi2007}. Consequently, the intensive moments (microscopic) converge to the macroscopic moments $\mu_k$
    \begin{equation*}
        M_k(t)\;\longrightarrow\;\mu_k(t) = \int x^kf(x, p, t)\dif{x}\dif{p}
    \end{equation*}
    as $N\to\infty$.
    
    By taking the limit $N \to \infty$ in the equations of motion in Eq.~\eqref{eq:eom2}, and replacing $M_k$ by $\mu_k$, we obtain the mean--field single--particle dynamics
    \begin{align*}
        \dot{x}_n = p_n,\quad\dot{p}_n = -\left[x_n + 2x_n^3 - 3x_n^2\mu_1(t) + 3x_n\mu_2(t) - \mu_3(t)\right].
    \end{align*}
    Since $\mu_k(t)$ does not depend on $n$, this leads to the decoupled set of ODEs
    \begin{equation}
        \label{eq:singleparticleEOM}
        \dot{x} = p_n,\quad\dot{p} = -\left[x + 2x^3 - 3x^2\mu_1(t) + 3x\mu_2(t) - \mu_3(t)\right].
    \end{equation}
    This equation describes the motion of a single ``mean--field particle'' in a self-consistent force field determined by the global distribution through the $\mu_k(t)$.

    In Hamiltonian dynamics, the probability density is transported along the flow, i.e., probability is conserved along trajectories (Liouville's theorem). Then the evolution of the probability distribution is governed by the Liouville equation:
    \begin{align*}
        \frac{\partial f}{\partial t} + \dot{x}\frac{\partial f}{\partial x} + \dot{p}\frac{\partial f}{\partial p} &= 0,
    \end{align*}
    which expresses the incompressibility of the Hamiltonian flow in phase space. From the equations of motion, with $\dot{p} = F(x, t)$, this equation takes the explicit form
    \begin{align*}
        \frac{\partial f}{\partial t} + p\frac{\partial f}{\partial x} + F(x, t)\frac{\partial f}{\partial p} &= 0,
    \end{align*}
    with the self-consistency equation
    \begin{equation}
        \label{eq:muk}
        \mu_k(t) = \int x^kf(x, p, t)\dif{x}\dif{p}.
    \end{equation}
    
    This is the Vlasov collisionless equation, where $F(x, t) = \dot{p}$ is the mean--field force. This means that the force $F$ depends on $f$, and $f$ evolves according to the force. For mean--field Hamiltonians of this type, the convergence of the microscopic dynamics to the Vlasov description has been established rigorously by Braun and Hepp~\cite{Braun1977} with the uniqueness of the solution later studied by Dobrushin~\cite{Dobrushin1979}. The Vlasov description has also been extensively studied in the context of Hamiltonian mean--field models~\cite{Latora1999, Antoniazzi2007}. 

    \subsection{Conditions for integrability}

    For the Vlasov equation, a sufficient condition for stationarity is that the distribution function depends on phase space coordinates only through a single--particle Hamiltonian. In particular, if the mean--field force is time independent and derives from an effective potential $V_{\mathrm{eff}}(x)$, the single--particle Hamiltonian can be written as
    \begin{equation}
        \label{eq:singleparticleham}
        h(x,p) = \frac{p^2}{2} + V_{\mathrm{eff}}(x).
    \end{equation}
    In this case, any distribution of the form $f(x,p) = \Phi(h)$ is a stationary solution of the Vlasov equation. This follows directly from the Hamiltonian structure of the Vlasov dynamics, since $\partial_t f + \{f,h\} = 0$ and $\{\Phi(h),h\} = 0$, where $\{g_1,g_2\}$ is the Poisson bracket of the two arbitrary functions of $x$, $p$, and $t$.
    
    For the quartic mean--field model considered here, if the macroscopic moments $\mu_k$ are time independent, the mean--field force
    \begin{equation*}
        F(x) = -\left(x + 2x^3 - 3x^2\mu_1 + 3x\mu_2 - \mu_3\right)
    \end{equation*}
    is conservative and can be written as $F(x) = -\dif V_{\mathrm{eff}}/\dif x$, with the effective potential
    \begin{equation*}
        V_{\mathrm{eff}}(x)
        =
        \frac{x^2}{2}
        + \frac{x^4}{2}
        - \mu_1 x^3
        + \frac{3}{2}\mu_2 x^2
        - \mu_3 x,
    \end{equation*}
    up to an irrelevant additive constant. The corresponding single--particle Hamiltonian is therefore explicitly time independent, and the microscopic dynamics reduces to an autonomous one-degree-of-freedom Hamiltonian system, which is integrable. Consequently, integrability of the Vlasov single--particle dynamics requires that the macroscopic moments $\mu_k$ be time independent.
    
    \begin{figure}[t]
        \centering
        \includegraphics[width=\linewidth]{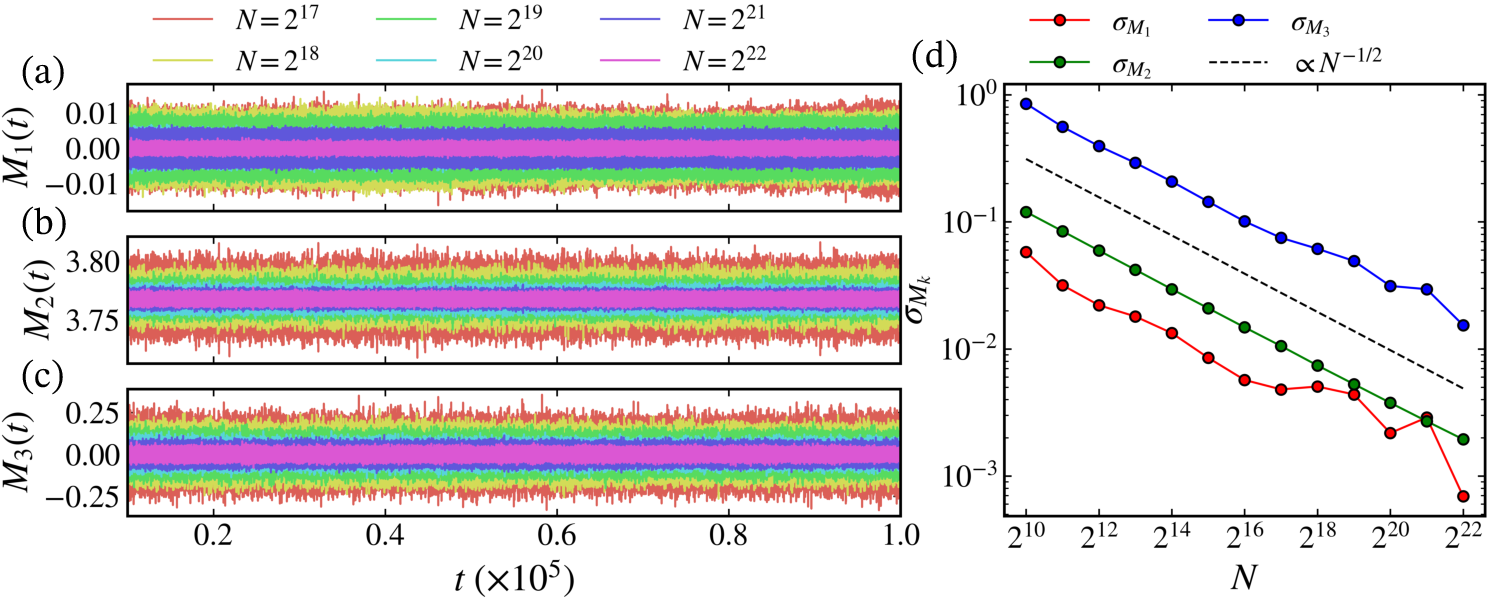}
        \caption{(a)--(c) Intensive moments $M_k(t)$ as a function of time for different values of $N$, and (d) the standard deviation $\sigma_{M_k}$ as a function of $N$, at $\varepsilon = 100$.}
        \label{fig:numerical_Mk}
    \end{figure}
    
    We therefore test the condition $\mu_k \equiv \mathrm{const}$ numerically by monitoring the finite--$N$ moments $M_k$ [Eq.~\eqref{eq:moments}] and verifying whether they converge to time-independent values as $N$ increases. In all numerical simulations, we adopt waterbag-type initial conditions: all coordinates are initially set to zero, while the momenta are independently drawn from a uniform distribution. The momenta are then rescaled to enforce zero total momentum, $\sum_n p_n(0)=0$, and subsequently rescaled to fix the total energy $E=\varepsilon N$, with $\varepsilon$ the specific energy.
    
    Panels (a)--(c) of Fig.~\ref{fig:numerical_Mk} show the time evolution of the intensive moments $M_k(t)$ for different $N$ at $\varepsilon=100$. Panel (d) shows the standard deviation $\sigma_{M_k}$ of $M_k(t)$ after an initial transient phase as a function of $N$, ranging from $N = 2^{10} = 1024$ to $N = 2^{22} = 4\,194\,304$. As $N$ increases, the amplitude of the $M_k$'s oscillations decreases, and $\sigma_{M_k}$ exhibits a clear scaling $\sigma_{M_k} \propto N^{-1/2}$. Therefore, in the thermodynamic limit the macroscopic moments $\mu_k$ become constant, the effective potential $V_{\mathrm{eff}}(x)$ is stationary, and the single--particle Hamiltonian in Eq.~\eqref{eq:singleparticleham} is time independent, confirming the integrability of the microscopic Vlasov dynamics. 
    
    At finite $N$, however, the moments fluctuate around their mean--field values,
    \begin{align*}
        M_k(t) = \mu_k + \delta M_k(t),
    \end{align*}
    where $\delta M_k$ arises from discreteness effects. Since $M_k$ is an average over $N$ particles, standard statistical arguments imply
    \begin{equation*}
        \delta M_k = \mathcal{O}(N^{-1/2}),
    \end{equation*}
    a scaling that is also verified numerically in Fig.~\ref{fig:numerical_Mk}(d). The finite--$N$ equations of motion can thus be written as
    \begin{equation*}
        \dot{x}_n = p_n,\quad\dot{p}_n = F_{\mathrm{mean-field}}(x_n(t)) + \mathcal{O}(N^{-1/2}),
    \end{equation*}
    i.e., the exact dynamics consists of mean--field motion perturbed by weak, stochastic-like fluctuations generated by the discrete particle nature of the system. Consequently,
    \begin{align*}
        \text{finite--$N$ dynamics}
        =
        \text{mean--field dynamics}
        + \text{weak noise of amplitude } N^{-1/2}.
    \end{align*}
    
    Since the Vlasov limit corresponds to integrable single--particle motion, the dominant source of chaos in the large (but finite) $N$ regime is the stochastic noise induced by these discreteness fluctuations~\cite{Casetti1993,Casetti1995,Casetti1996}, which vanishes as $N\rightarrow\infty$. Therefore, in the thermodynamic limit the largest Lyapunov exponent satisfies
    \begin{equation*}
        \lambda_1(N\to\infty)=0.
    \end{equation*}
    At finite $N$, the stochastic perturbation $\delta M_k(t)$ breaks integrability and generates chaos with a nonzero Lyapunov exponent. Hence, the origin of chaos in this model is purely \textit{granular}~\cite{Casetti1993,Casetti1995,Casetti1996}, arising from finite--$N$ fluctuations around the Vlasov mean field.

    \subsection{Integral of motion}
    \label{subsec:integralsofmotion}
    
    We have shown that in the thermodynamic limit the dynamics becomes effectively integrable, provided the macroscopic moments $\mu_k$ become time independent. In this limit, the equations of motion decouple into a set of independent ordinary differential equations [Eq.~\eqref{eq:singleparticleEOM}], each describing the motion of a single mean--field particle governed by an autonomous one-degree-of-freedom Hamiltonian [Eq.~\eqref{eq:singleparticleham}]. The corresponding single--particle energy,
    \begin{equation}
        \label{eq:Eeff}
        h(x,p)
        =
        \frac{p^2}{2}
        + \frac{x^2}{2}
        + \frac{x^4}{2}
        - \mu_1 x^3
        + \frac{3}{2}\mu_2 x^2
        - \mu_3 x
        \equiv E_{\mathrm{eff}},
    \end{equation}
    is then an integral of motion.
    
    A complementary numerical test of integrability in the thermodynamic limit is therefore to examine whether the effective energies $E_{\mathrm{eff},n}(t)$ associated with each site become time independent as $N$ increases. If integrability is recovered only asymptotically, these quantities need not be conserved at finite $N$, but their fluctuations must vanish in the limit $N\to\infty$. To quantify this behavior, we compute the average standard deviation of the effective energies,
    \begin{equation}
        \label{eq:stds}
        \sigma(N)
        =
        \frac{1}{N}
        \sum_{n=1}^N
        \sqrt{
            \frac{1}{T}
            \sum_{t=0}^{T}
            \left[
                E_{\mathrm{eff},n}(t)
                -
                \overline{E}_{\mathrm{eff},n}
            \right]^2
        },
    \end{equation}
    where
    \begin{equation*}
        \overline{E}_{\mathrm{eff},n}
        =
        \frac{1}{T}
        \sum_{t=0}^{T}
        E_{\mathrm{eff},n}(t),
    \end{equation*}
    using a total integration time $T = 1.0\times10^5$.
    
\begin{figure}[t]
        \centering
        \includegraphics[width=\linewidth]{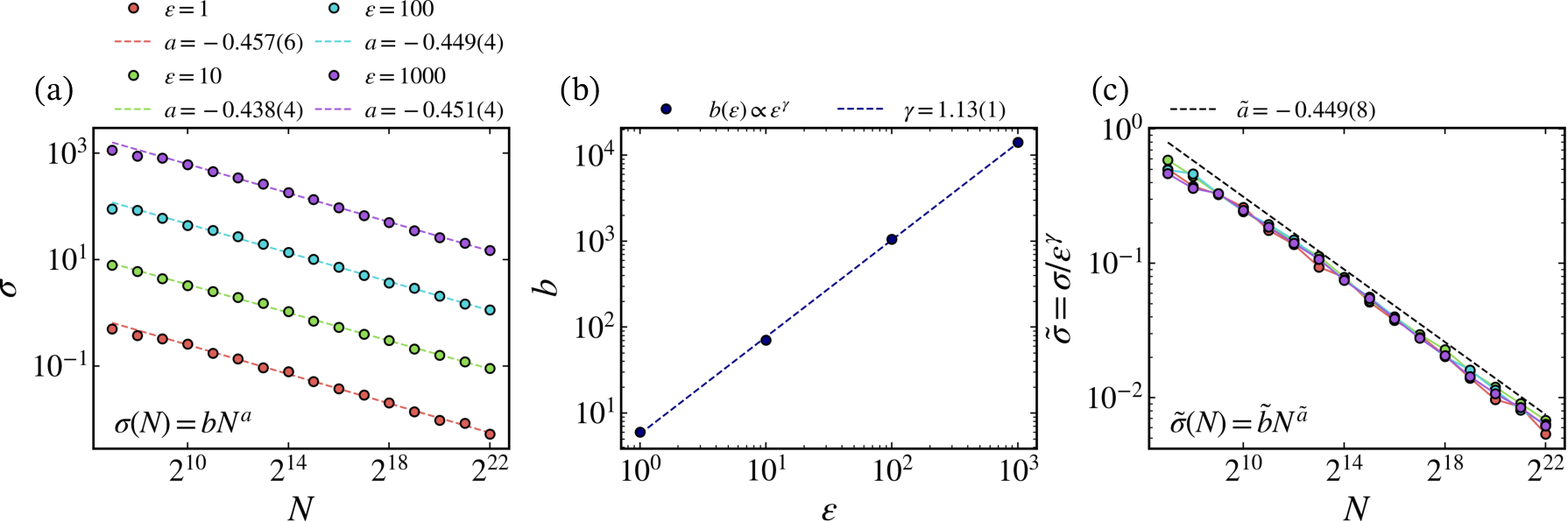}
        \caption{
        (a) Average standard deviation of the effective energies [Eq.~\eqref{eq:stds}] as a function of the number of degrees of freedom $N$ for four values of the specific energy $\varepsilon$.
        The data follow a power-law decay $\sigma(N)=bN^{a}$.
        (b) Dependence of the prefactor $b$ on $\varepsilon$, showing a power-law scaling $b(\varepsilon)\propto\varepsilon^{\gamma}$.
        (c) Collapse of the curves in panel (a) after rescaling $\tilde\sigma=\sigma/\varepsilon^{\gamma}$, indicating scale invariance of the effective-energy fluctuations with respect to $\varepsilon$.
        }
        \label{fig:std}
    \end{figure}

    Figure~\ref{fig:std}(a) shows $\sigma(N)$ for $\varepsilon=1,10,100,$ and $1000$. In all cases, the fluctuations of the effective energies decay algebraically with $N$, following $\sigma(N)=bN^{a}$. This decay demonstrates that, although $E_{\mathrm{eff},n}(t)$ is not conserved at finite $N$, its fluctuations vanish in the thermodynamic limit, confirming that Eq.~\eqref{eq:Eeff} becomes an exact integral of motion only as $N\to\infty$.
    
    The fitted power-law scaling also allows us to estimate the system size $N^*$ beyond which effective-energy conservation would become observable within double-precision numerical accuracy, defined by $\sigma(N^*)<10^{-16}$. Using the fitted values of $a$ and $b$, we find $N^*\sim10^{36},10^{40},10^{42}$, and $10^{44}$ for $\varepsilon=1,10,100,$ and $1000$, respectively. These values are far beyond any feasible numerical simulation, explaining why strict conservation of $E_{\mathrm{eff},n}$ cannot be observed in practice, despite being guaranteed in the thermodynamic limit.

    As shown in Fig.~\ref{fig:std}(a), the decay exponent $a$ is approximately the same for all $\varepsilon$, while the curves differ only by a multiplicative prefactor. Figure~\ref{fig:std}(b) reveals that this prefactor scales with the specific energy as $b(\varepsilon)\propto\varepsilon^{\gamma}$. This scaling allows for a rescaling of the fluctuations, $\tilde\sigma=\sigma/\varepsilon^{\gamma}$, leading to the curves collapsing, shown in Fig.~\ref{fig:std}(c). The collapse indicates that the fluctuations of the effective on-site energy are scale invariant with respect to $\varepsilon$, exhibiting universal statistical behavior across different energy scales~\cite{leonelscaling,Leonel2004,Leonel2007}.

    In summary, stationarity of the Vlasov description in the thermodynamic limit corresponds to the macroscopic moments $\mu_k$ becoming time independent, which makes the mean--field force and the effective single--particle Hamiltonian $h(x,p)$ autonomous. Numerically, this regime is identified by the vanishing of the temporal fluctuations of the intensive moments $M_k(t)$ and by the progressive conservation of the effective single--particle energies $E_{\mathrm{eff}}$ as $N\to\infty$. In Sec.~\ref{sec:finiteNdynamics}, we turn to the dynamical and statistical properties of the model at finite $N$. We show numerically that the largest Lyapunov exponent decreases with increasing system size and vanishes in the thermodynamic limit, and we analyze the resulting finite--$N$ dynamics within the framework of non--extensive statistical mechanics~\cite{Tsallis2009}.

    \section{The dynamics and statistics for finite $N$}
    \label{sec:finiteNdynamics}

    In this section, we investigate the dynamical behavior of the mean--field Hamiltonian defined in Eq.~\eqref{eq:hamiltonian} for finite values of $N$. Our primary goal is to characterize how chaos evolves as the number of degrees of freedom increases. To this end, we compute the largest Lyapunov exponent (LLE) for progressively larger $N$ and analyze the resulting dynamical regimes. We interpret these regimes using concepts from non--extensive statistical mechanics~\cite{Tsallis2009}.
    
    The numerical computation of Lyapunov exponents requires the simultaneous integration of the equations of motion and the associated variational equations~\cite{Benettin1980, Wolf1985}. For Hamiltonian systems, it is essential that the variational integration also be performed using a symplectic scheme~\cite{Skokos2010}. We now briefly outline the formulation of the variational equations for Hamiltonian systems, which will be used in the computation of the largest Lyapunov exponent.

    Let $\mathbf{y} = (\mathbf{x}, \mathbf{p})^T$ be the state vector in the $2N$-dimensional phase space. The time evolution of this vector is given by Hamilton's equations, which in matrix form, are
    \begin{equation*}
        \frac{\dif{\mathbf{y}}}{\dif{t}} =
        \begin{pmatrix}
            \displaystyle \frac{\partial H}{\partial \mathbf{p}} &
            \displaystyle -\frac{\partial H}{\partial \mathbf{x}}
        \end{pmatrix}^T = \mathbf{J}_{2N}\cdot \mathbf{D}_H(\mathbf{x}(t)),
    \end{equation*}
    where
    \begin{equation*}
        \mathbf{J}_{2N} = \begin{pmatrix}
            \mathbf{0}_N & \mathbf{I}_N\\
            -\mathbf{I}_N & \mathbf{0}_N
        \end{pmatrix},
    \end{equation*}
    with $\mathbf{I}_N$ being the $N\times N$ identity matrix and $\mathbf{0}_N$ a $N\times N$ matrix of zeros, and 
    \begin{equation*}
        \mathbf{D}_H(\mathbf{x}(t)) = \begin{pmatrix}
            \displaystyle \frac{\partial H}{\partial x_1} & \displaystyle \frac{\partial H}{\partial x_2} & \cdots & \displaystyle \frac{\partial H}{\partial x_N} & \displaystyle \frac{\partial H}{\partial p_1} & \displaystyle \frac{\partial H}{\partial p_2} & \cdots & \displaystyle \frac{\partial H}{\partial p_N}
        \end{pmatrix}^T
    \end{equation*}
    is the gradient of the Hamiltonian evaluated at $\mathbf{x}(t)$. The time evolution of a deviation vector $\mathbf{w} = (\delta\mathbf{x}, \delta\mathbf{p})^T$ is governed by the variational equations:
    \begin{equation}
        \label{eq:vareq}
        \frac{\dif{\mathbf{w}}}{\dif{t}} = \left[\mathbf{J}_{2N}\cdot \mathbf{D}^2_H\left(\mathbf{x}(t)\right)\right]\cdot\mathbf{w},
    \end{equation}
    where $\mathbf{D}^2_H$ is the Hessian of the Hamiltonian evaluated at $\mathbf{x}(t)$:
    \begin{equation*}
        \mathbf{D}^2_H =
        \begin{pmatrix}
            \displaystyle \frac{\partial^2 H}{\partial \mathbf{x}^2}
            &
            \displaystyle \frac{\partial^2 H}{\partial \mathbf{x}\,\partial \mathbf{p}}
            \\
            \displaystyle \frac{\partial^2 H}{\partial \mathbf{p}\,\partial \mathbf{x}}
            &
            \displaystyle \frac{\partial^2 H}{\partial \mathbf{p}^2}
        \end{pmatrix} = \begin{pmatrix}
            \mathbf{H}_{xx} & \mathbf{H}_{xp} \\
            \mathbf{H}_{px} & \mathbf{H}_{pp}
        \end{pmatrix}.
    \end{equation*}
    For a separable Hamiltonian function, i.e., $H = T(\mathbf{p}) + V(\mathbf{x})$, such as ours in Eq.~\eqref{eq:hamiltonian}, $\mathbf{H}_{xp} = \mathbf{H}_{px} = 0$, and the variational equations, given by Eq.~\eqref{eq:vareq}, can be explicitly written as
    \begin{equation*}
        \begin{aligned}
            \frac{\dif{\delta\mathbf{x}}}{\dif{t}} &= \mathbf{H}_{pp}\cdot\delta\mathbf{p},\\
            \frac{\dif{\delta\mathbf{p}}}{\dif{t}}  &= -\mathbf{H}_{xx}\cdot\delta\mathbf{x}.\\
        \end{aligned}
    \end{equation*}
    The LLE can then be calculated as
    \begin{equation*}
        \lambda_1
        = \lim_{t\to\infty}
        \frac{1}{t}
        \log\frac{\|\mathbf{w}(t)\|}{\|\mathbf{w}(0)\|}
        \approx
        \frac{1}{T}
        \sum_{n=1}^{n_{\mathrm{steps}}}
        \log \|\mathbf{w}(t_n)\|,
    \end{equation*}
    where the first expression is the theoretical definition and the second one represents the numerical implementation. Here $t_n = n\Delta t$, with $\Delta t$ the integration time step and $T = n_{\mathrm{steps}}\Delta t$ the total integration time. The deviation vector $\mathbf{w}(t)$ is renormalized to unit norm at each step and only the instantaneous growth factor $\|\mathbf{w}(t_n)\|$ contributes to the sum. For chaotic trajectories, $\|\mathbf{w}(t)\|$ grows exponentially, making
    periodic renormalization essential to avoid numerical overflow~\cite{Benettin1980}.

    \begin{figure}[t]
        \centering
        \includegraphics[width=\linewidth]{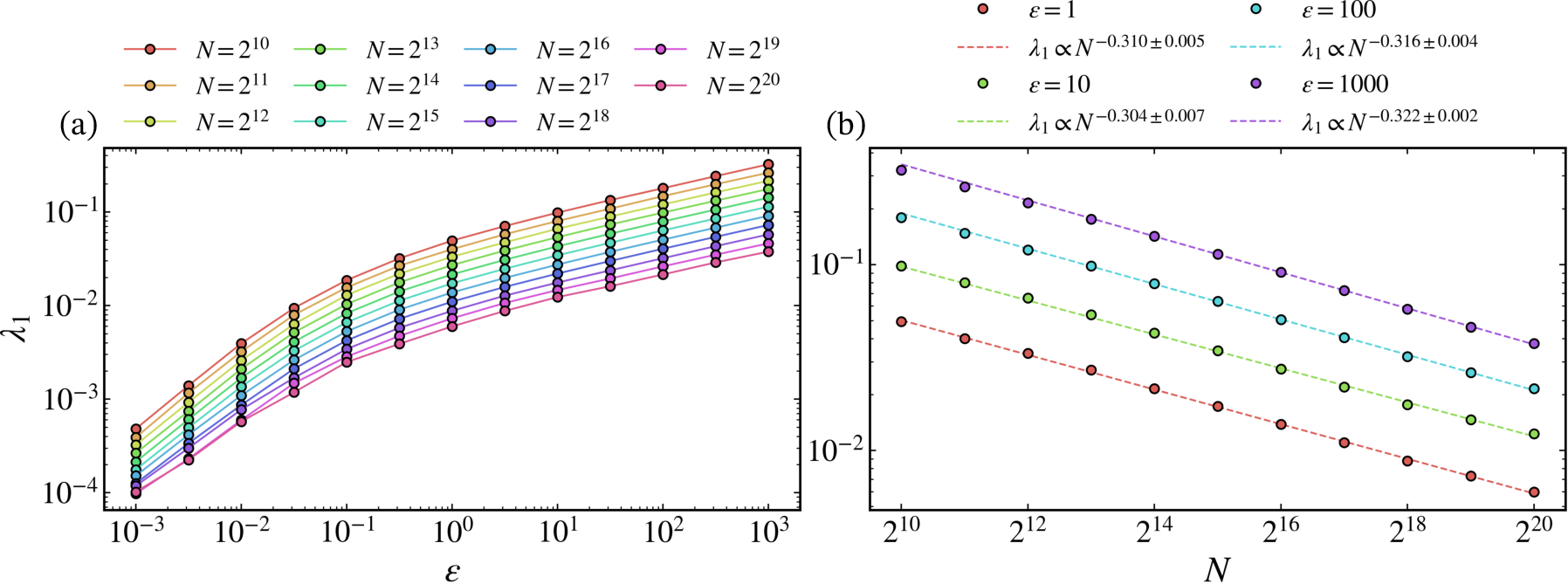}
        \caption{The largest Lyapunov exponent (a) as a function of the specific energy $\varepsilon$ for different numbers of degrees of freedom $N$ and (b) as a function of the number of degrees of freedom $N$ for different specific energies $\varepsilon$.}
        \label{fig:LLE}
    \end{figure}

     In Fig.~\ref{fig:LLE}, we show the behavior of the LLE, $\lambda_1$. In panel (a), $\lambda_1$ is shown as a function of the specific energy $\varepsilon$ for several $N$, while panel (b) displays $\lambda_1$ as a function of $N$ for fixed $\varepsilon$. Our simulations span an exceptionally wide range of parameters, with the number of degree of freedom ranging from $N = 2^{10} = 1024$ up to $N = 2^{20}=1\,048\,576$, and specific energies covering six decades, $10^{-3}\leq\varepsilon\leq10^{3}$. Figure~\ref{fig:LLE}(a) shows that $\lambda_1$ increases monotonically with $\varepsilon$ for all $N$, reflecting the progressive strengthening of chaotic dynamics as the specific energy grows. 

    Additionally, Fig.~\ref{fig:LLE}(b) demonstrates that, for all values of $\varepsilon$, the largest Lyapunov exponent decays algebraically with the number of degrees of freedom, $\lambda_1 \propto N^{-\alpha}$, with a robust exponent $\alpha \simeq 0.31$. While similar trends for the quartic Hamiltonian have been reported previously, most recently in Ref.~\cite{Christodoulidi2025} for relatively small $N$ ($N_{\mathrm{max}} = 2^{15}$), our results extend this scaling over more than two orders of magnitude in $N$, providing strong numerical evidence that this behavior persists deep into the thermodynamic limit. The decay of the largest Lyapunov exponent with increasing $N$ is a well-known phenomenon in fully coupled and long--ranged Hamiltonian systems~\cite{Anteneodo1998, Latora1998, Dauxois2002, Ginelli2011, Manos2011, Christodoulidi2014, Christodoulidi2016, Christodoulidi2018, Christodoulidi2025}. In particular, the asymptotic scaling $\lambda_1 \propto N^{-1/3}$ has been derived analytically for specific models~\cite{Firpo1998, Latora1999, Firpo2001, Anteneodo2001}. Our measured exponent is fully consistent with this theoretical prediction within numerical accuracy, and, crucially, its persistence over such large $N$ and energy scales reinforces the view that this scaling is a universal dynamical signature of fully coupled and long--ranged Hamiltonian dynamics.

    Having established the scaling behavior of the LLE, we now turn to the characterization of the dynamical regime, namely whether the system exhibits week or strong rchaos. In Ref.~\cite{Christodoulidi2025}, the observed decay of $\lambda_1$ with $N$ was interpreted as indicative of a weakly chaotic dynamics. However, the vanishing of the LLE in the thermodynamic limit is not, by itself, sufficient to classify the nature of chaos for finite $N$. To address this issue, we characterize the dynamics within the formalism of non--extensive statistical mechanics, which has proven effective in distinguishing between strong and weak chaos in both symplectic maps~\cite{Gerakopoulos2008, Antonopoulos2016, Tirnakli2016, Ruiz2017, Ruiz2017b, Tirnakli2020, Bountis2023, Zulkarnain2026} and Hamiltonian systems~\cite{Leo2010, Antonopoulos2011, Antonopoulos2014, Christodoulidi2016, Antonopoulos2011b}. Specifically, we analyze the temporal evolution of the entropic index $q$ associated with $q$-generalized statistics. The value of $q$ provides a sensitive diagnostic of the degree of chaoticity: convergence to $q = 1$ signals strong (Boltzmann–Gibbs) chaos, whereas persistent deviations from unity are commonly associated with weak chaos and long--lived dynamical correlations~\cite{Moyano2006, Antonopoulos2011b, Bountis2012}. The entropic index $q$ is associated with the $q$--Gaussian probability density function (PDF)~\cite{Tsallis2009}
    \begin{equation}
        \label{eq:qgaussian}
        P(x; q) = \alpha\left[1 + \beta \left(q - 1\right)x^2\right]^{\frac{1}{1 - q}},
    \end{equation}
    where $\alpha$ is a normalization constant and $\beta$ an arbitrary parameter. Equation~\eqref{eq:qgaussian} is the generalization of the Gaussian PDF and, in the limit $q\to 1$, $P(x; 1)$ becomes the Gaussian PDF.
    
    In the following, we characterize the dynamical behavior by computing the probability density function (PDF) of the momenta and estimating the time-dependent entropic index $q(t)$. As a probe of the dynamics, we use all of the momenta $p_n(t)$. For a given specific energy $\varepsilon$, we define a symmetric momentum interval between $-p^\star$ and $p^\star$, with $p^\star = 7.5\sqrt{2\varepsilon}$, and discretize this interval into 1000 bins. This choice of $p^\star$ guarantees that, during the entire temporal evolution, all values of $p_n(t)$ remain within this interval for each $\varepsilon$. After discarding an initial transient of $10^4$ units of time, we sample the momenta every 5 time steps. At each sampling time, we count how many $p_n(t)$ fall into each bin, and accumulate these counts over time to construct a cumulative PDF. The total sampling time is $T = 1.0\times10^{10}$. Finally, we compute the normalized PDF $P(p)/P(0)$ as a function of $p P(0)$, and fit this distribution to estimate the entropic index $q(t)$.

    \begin{figure}[t]
        \centering
        \includegraphics[width=\linewidth]{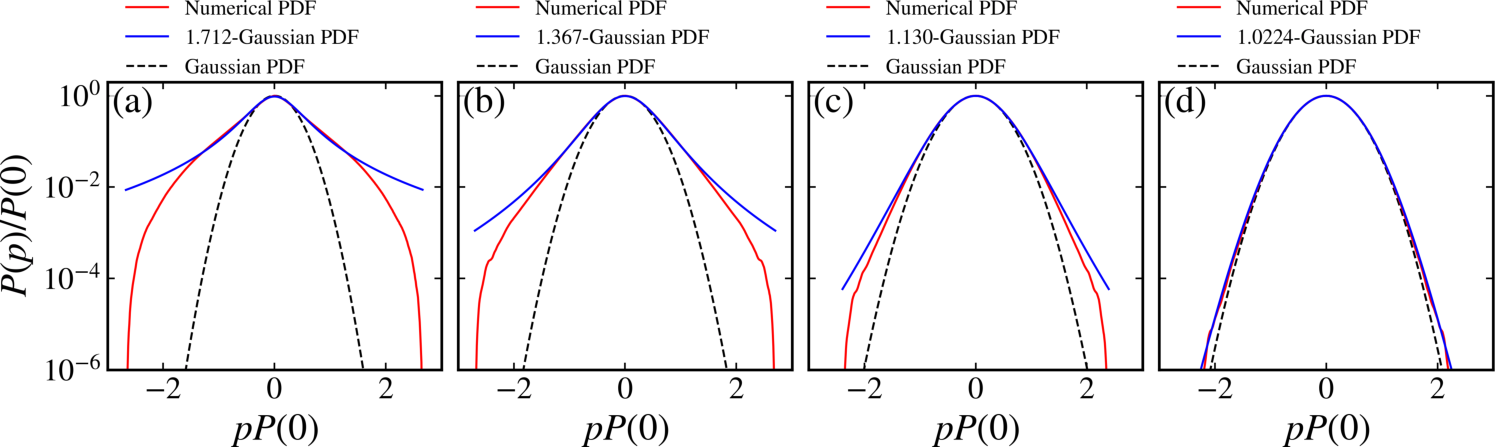}
        \caption{(red curves) Numerical probability density functions (PDFs) for $N = 1024$ and $\varepsilon = 100$ evaluated at different times: (a) $t = 1.0\times10^{5}$, (b) $t = 1.0\times10^{6}$, (c) $t = 1.0\times10^{7}$, and (d) $t = 1.0\times10^{8}$. The blue curves show the best fits to the numerical PDFs, yielding (a) $q = 1.712\pm0.008$, (b) $q = 1.367\pm0.002$, (c) $q = 1.130\pm0.001$, and (d) $q = 1.0224\pm0.0003$. The dashed black curves represent the Gaussian PDFs ($q = 1$) shown for comparison. Note that as $t \to \infty$, $q \to 1$. Table~\ref{tab:qvalues} reports two quantitative measures of the fit quality for the distributions shown in this figure, namely the mean squared error (MSE) and the mean absolute error (MAE).}
        \label{fig:pdfs}
    \end{figure}

    \begin{table}[b!]
        \centering
        \caption{Estimated values of the entropic index $q$ obtained from the fitting procedure of PDFs in Fig.~\ref{fig:pdfs} for different instants of time $t$. The reported uncertainty $\sigma_q$ corresponds to the fitting error. The quality of each fit is quantified by the mean squared error (MSE) and mean absolute error (MAE), both of which decrease systematically as $t$ increases, indicating an improved agreement between the fitted model and the numerical data.}
        \label{tab:qvalues}
        \begin{tabular}{ccccc}
            \toprule
            Fig. & $t$ & $q \pm \sigma_q$ & MSE & MAE \\
            \midrule
            \ref{fig:pdfs}(a) & $1.0\times10^5$ & $1.712 \pm 0.008$ & $1.18\times10^{-4}$ & $9.79\times10^{-3}$ \\
            \ref{fig:pdfs}(b) & $1.0\times10^6$ & $1.367 \pm 0.002$ & $1.20\times10^{-5}$ & $2.93\times10^{-3}$ \\
            \ref{fig:pdfs}(c) & $1.0\times10^7$ & $1.130 \pm 0.001$ & $3.85\times10^{-6}$ & $1.52\times10^{-3}$ \\
            \ref{fig:pdfs}(d) & $1.0\times10^8$ & $1.0224 \pm 0.0003$ & $1.93\times10^{-7}$ & $3.25\times10^{-4}$ \\
            \bottomrule
        \end{tabular}
    \end{table}

    Figure~\ref{fig:pdfs} shows examples of these normalized PDFs for $N = 1024$ and $\varepsilon = 100$ (red curves), together with the fitted $q$--Gaussians (blue curves) and the standard Gaussian (dashed black curves) for four different instants of time. In Table~\ref{tab:qvalues}, we report the fitted $q$ values and their associated standard errors, obtained from the covariance matrix of the fit, as well as two quantities that quantify the quality of each fit, namely the mean squared error (MSE) and the mean absolute error (MAE). The MSE measures the average squared deviation between the data and the fitted values, while the MAE measures the average absolute deviation between the data and the fitted values. For a dataset $y_i$ and corresponding fitted values $\hat{y}_i$, with $i = 1, 2, \ldots, n$, these quantities are defined as
    \begin{equation*}
        \mathrm{MSE} = \frac{1}{n}\sum_{i=1}^n (y_i - \hat{y}_i)^2,\quad\mathrm{MAE} = \frac{1}{n}\sum_{i=1}^n |y_i - \hat{y}_i|.
    \end{equation*}

    Table~\ref{tab:qvalues} quantifies the time dependence of the entropic index $q$ obtained from the PDFs discussed above. While values $q > 1$ are commonly associated with weak chaos and long--lived correlations~\cite{Moyano2006, Antonopoulos2011b, Bountis2012}, it is important to note that deviations from $q = 1$ may also arise from finite-time effects and incomplete statistical sampling, particularly at early stages of the evolution~\cite{Gerakopoulos2008,Ruiz2017,Ruiz2017b,Tirnakli2016,Zulkarnain2026}. At short times ($t = 10^5$), the large value $q = 1.712$ coincides with relatively large values of both the MSE and MAE, indicating a poorer overall quality of the fit. This suggests that the observed deviation from Gaussian statistics at early times, at least in part, reflect insufficient sampling of the momenta rather than an intrinsic non-Boltzmann–Gibbs regime. As time increases, both error measures decrease systematically by several orders of magnitude, signaling that the PDFs become progressively smoother and better resolved. 

    Additionally, the fitted values of $q$ decrease monotonically and approach unity, reaching $q = 1.0224$ at $t = 10^8$, where the MSE and MAE exhibit their smallest values. This simultaneous convergence of $q \to 1$ and of the fit errors towards very small values supports the interpretation that the late-time PDFs are genuinely Gaussian. Therefore, our results support the interpretation that the dynamics is \textit{strongly chaotic}: although values $q > 1$ are observed at short times, these deviations can be attributed to finite-time and finite-sampling effects rather than to an intrinsically weakly chaotic regime. As the observation time increases, the entropic index $q$ systematically approaches unity, and the corresponding PDFs converge to the Gaussian form, consistent with Boltzmann–Gibbs statistics. Within this framework, the transient deviations of $q$ from unity reflect the finite-time nature of the measurements, while the asymptotic behavior provides clear evidence for strong chaos. In this context, it is worth noting that the $q > 1$ values reported in Ref.~\cite{Christodoulidi2025} were obtained using a maximum sampling time of order $t = 10^{5}$. As shown in Table~\ref{tab:qvalues}, this time scale corresponds precisely to the early-time regime in which finite-time and finite-sampling effects are still dominant and the PDFs are not yet fully converged. Our results therefore indicate that the previously reported $q > 1$ values arise from the limited observation time rather than from a genuinely weakly chaotic or non-Boltzmann-Gibbs regime.

    \begin{figure}[t]
        \centering
        \includegraphics[width=\linewidth]{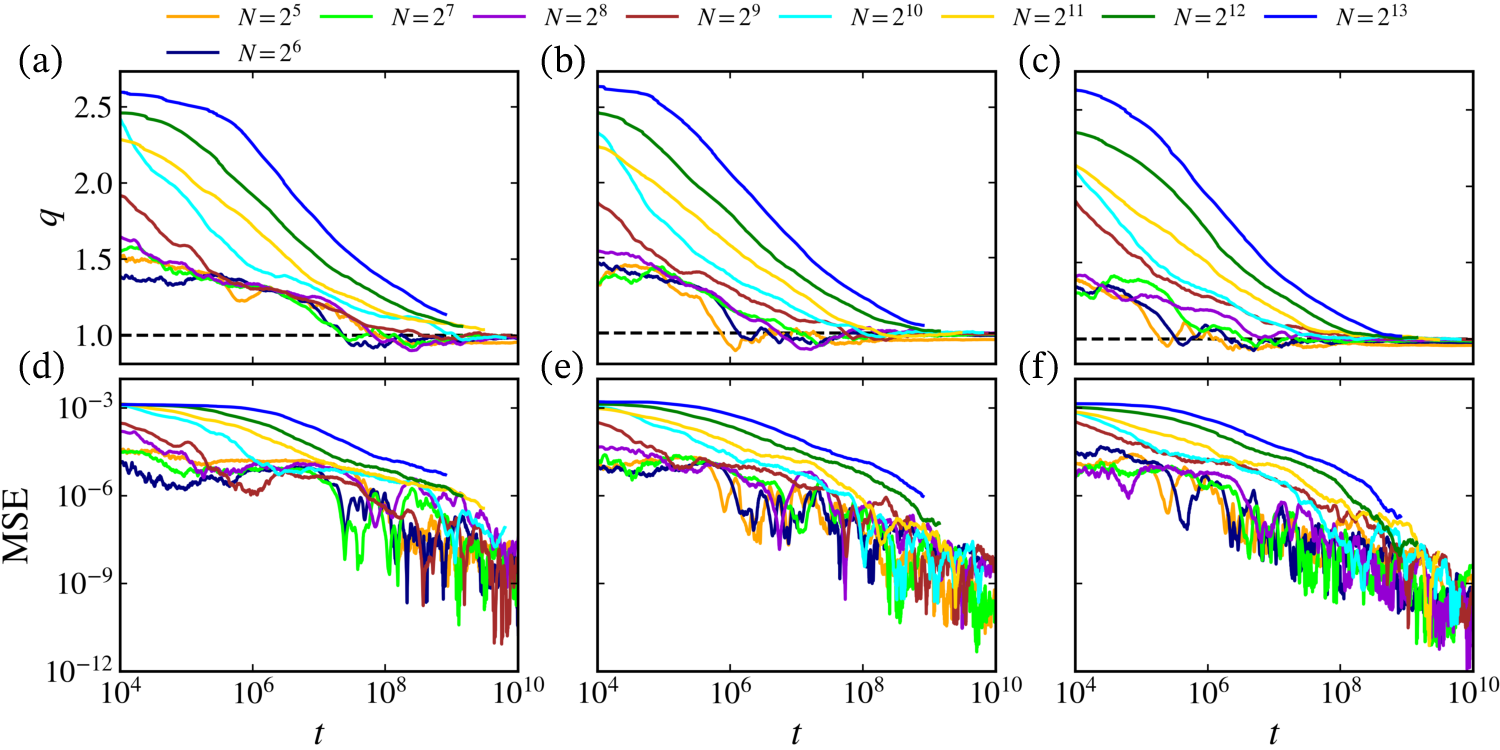}
        \caption{(a)--(c) Entropic index $q$ and (d)--(f) the mean squared error (MSE) as a function of time for different number of degrees of freedom $N$ and different specific energy $\varepsilon$: (a) and (d) $\varepsilon = 10$, (b) and (e) $\varepsilon = 100$, and (c) and (f) $\varepsilon = 1000$. The $q$ values have been obtained via the best fit of the numerical PDFs at each instant of time.}
        \label{fig:q_history}
    \end{figure}

    Figure~\ref{fig:q_history} further corroborates the interpretation drawn from the PDFs in Fig.~\ref{fig:pdfs} and Table~\ref{tab:qvalues} by displaying the full temporal evolution of the entropic index $q(t)$ [Figs.~\ref{fig:q_history}(a)--\ref{fig:q_history}(c)] together with the corresponding mean squared error (MSE) of the fits [Figs.~\ref{fig:q_history}(d)--\ref{fig:q_history}(f)] over six decades in time, for a wide range of $N$ and $\varepsilon$. Panels (a)--(c) show $q(t)$ and the associated MSE in panels (d)--(f) for $\varepsilon = 10, 100$, and $1000$, respectively, and for the number of degrees of freedom ranging from $N = 2^5 = 32$ up to $N = 2^{13} = 8\,192$. In all cases, $q(t)$ decreases monotonically and approaches unity at long times, while the MSE also decreases by several orders of magnitude. This simultaneous convergence provides strong evidence that the asymptotic dynamics is strongly chaotic. In this context, values $q(t) > 1$ observed at intermediate times should be interpreted as finite-time effects associated with incomplete statistical sampling rather than as signatures of a weakly chaotic regime. This interpretation is further supported by the MSE evolution: deviations of $q$ from unity systematically coincide with relatively large MSE values, whereas in the long-time regime the MSE reaches extremely small values ($\sim 10^{-10}$), indicating well-resolved Gaussian statistics consistent with Boltzmann–Gibbs behavior.

    The conclusion that $\lambda_1 \to 0$ as $N \to \infty$ provided in Secs.~\ref{subsec:vlasovlimit} and \ref{subsec:integralsofmotion} and in Fig.~\ref{fig:LLE}(b) and that $q \to 1$ for all $N$ seems to be in seems to be in contradiction. However, the key point is that the LLE only becomes zero in the strict thermodynamic limit. For any large but finite $N$, chaos originates from parametric instability, i.e., the stochastic noise that decreases with $N$ as $\mathcal{O}(N^{-1/2})$~\cite{Casetti1993, Casetti1995, Casetti1996}. This stochasticity is the mechanism that seeds chaos at finite $N$, and therefore, for large finite $N$, the system is fully chaotic, with no surviving invariant tori and no regular regions. As $N$ increases, the amplitude of the noise decreases as $N^{-1/2}$ and the LLE decreases accordingly. Chaos, in this case, becomes ``weaker'' (slower divergence, smaller $\lambda_1$), but \textit{it remains strong chaos in the sense that the phase space contains no regular regions, only a uniform chaotic component}. 
    
    Additionally, Figs.~\ref{fig:q_history}(a)--\ref{fig:q_history}(c) shows that the convergence of $q(t)\to 1$ becomes progressively slower as $N$ increases. At first glance, this behavior could suggest the presence of long--lived non--Gaussian states~\cite{Moyano2006, Antonopoulos2011b, Bountis2012}. However, it can be consistently understood within the same dynamical picture discussed above. Chaos at finite $N$ is seeded by an effective stochastic component whose amplitude decreases with the number of degrees of freedom. As $N$ grows, this stochasticity becomes weaker and, consequently, longer times are required for statistical observables, such as the momentum PDFs, to approach their asymptotic Gaussian form. In this sense, the delayed convergence of $q(t)$ reflects the diminishing strength of chaotic fluctuations with increasing $N$, rather than the presence of a distinct long--lived non-Boltzmann-Gibbs regime. Moreover, the fact that $q(t)$ \emph{monotonically} converges to unity at long times is incompatible with the existence of long--lived non-Boltzmann-Gibbs states and supports the interpretation that all deviations from Gaussian statistics are transient.

    \section{Conclusions}
    \label{sec:concl}

    In this paper, we have investigated the dynamical and statistical properties of a quartic mean--field Hamiltonian model. We first addressed the thermodynamic limit of the model within the Vlasov collisionless framework. In this limit, the dynamics is effectively described by an autonomous one-degree-of-freedom Hamiltonian system, implying integrability of the microscopic motion. We verified this property numerically by analyzing the fluctuations of the intensive moments $M_k$ and of the single--particle effective energy $E_{\mathrm{eff}}$ as functions of the number of degrees of freedom $N$. We showed that the corresponding standard deviations decay algebraically with system size, following $\sigma \propto N^{-1/2}$, demonstrating that the mean--field force is indeed time independent, so that the single--particle Hamiltonian is autonomous and the single--particle dynamics is integrable.  Additionally, the single--particle effective energy becomes a constant of motion only in the strict thermodynamic limit. This provides direct numerical evidence that the system is integrable, asymptotically, as $N \to \infty$.

    We then focused on the dynamical behavior of the model for finite, but large, values of $N$. We computed the largest Lyapunov exponent (LLE) over an exceptionally wide range of $N$, from $N = 2^{10} = 1024$ up to $N = 2^{20} = 1\,048\,576$. Our results show that the LLE decays algebraically with $N$ according to $\lambda_1 \propto N^{-0.31}$, extending previous numerical studies of the same model by more than two orders of magnitude in $N$~\cite{Christodoulidi2025}. This scaling is fully consistent with the theoretical and numerical predictions $\lambda_1 \propto N^{-1/3}$ reported for several long--ranged and mean--field Hamiltonian systems~\cite{Firpo1998,Latora1999,Firpo2001,Anteneodo2001}.

    In Ref.~\cite{Christodoulidi2025}, the observed decay of $\lambda_1$ with $N$ was interpreted as evidence of weakly chaotic dynamics. To clarify this point, we employed tools from non--extensive statistical mechanics and analyzed the time evolution of the entropic index $q$ for different values of $N$. We found that, in all cases, $q$ converges monotonically to unity as the sampling time increases. This behavior is characteristic of strongly chaotic dynamics and rules out the presence of long--lived non-Boltzmann-Gibbs states, which are typically associated with values $q>1$~\cite{Moyano2006,Antonopoulos2011b,Bountis2012}. We therefore conclude that the $q>1$ values reported in Ref.~\cite{Christodoulidi2025} originate from finite-time effects rather than from a genuinely weakly chaotic regime. As the observation time is increased, these transient deviations disappear and the system ultimately exhibits fully developed chaotic dynamics at finite $N$.

    \section*{Code availability}

    The source code to reproduce the results reported in this paper is freely available in the \href{https://github.com/mrolims-publications/quartic-mean-field-hamiltonian.git}{GitHub repository} in Ref.~\cite{github}.

    \section*{Acknowledgments}

    This work was supported by the São Paulo Research Foundation (FAPESP, Brazil), under Grant Nos.~2019/14038-6, 2021/09519-5, 2023/08698-9 and 2024/09208-8, and by the National Council for Scientific and Technological Development (CNPq, Brazil), under Grant Nos.~301318/2019-0 and 304398/2023-3. The computational simulations in this research was supported by resources supplied by the Center for Scientific Computing (NCC/GridUNESP) of the São Paulo State University (UNESP), by the Oscillation Control Group of the University of São Paulo, and by the High Performance Computing Facility (Ceres) and associated support services at the University of Essex.


%

  \end{document}